\documentclass[osajnl,twocolumn,showpacs]{revtex4}
\usepackage{graphicx}
\newcommand{\be}{\begin{equation}}
\newcommand{\ee}{\end{equation}}

\begin{document}

\title{Nonlinear localized modes at phase-slip defects in waveguide arrays}

\author{Mario I. Molina$^{1,2}$ and Yuri S. Kivshar$^2$}

\affiliation{$^1$Departmento de F\'{\i}sica, Facultad de Ciencias,
Universidad de Chile, Santiago, Chile\\
$^2$Nonlinear Physics Center, Research School of Physical Sciences
and Engineering, Australian National University, Canberra ACT 0200,
Australia}

\begin{abstract}
We study light localization at a phase-slip defect created by two
semi-infinite mismatched identical arrays of coupled optical
waveguides. We demonstrate that the nonlinear defect modes possess
the specific properties of both nonlinear surface modes and
discrete solitons. We analyze stability of the localized modes and
their generation in both linear and nonlinear regimes.
\end{abstract}

\ocis{060.4370;190.4350;190;6135}

\maketitle

\newpage

The study of nonlinear dynamics in discrete systems has attracted
a special attention recently due to novel physics and possible
interesting applications~\cite{review}. In particular, it is well
known that discrete photonic systems can support different types
of spatially localized states in the form of discrete
solitons~\cite{review,book}. These solitons can be controlled by
the insertion of suitable defects in an array, as was suggested
theoretically~\cite{krolik,aceves} and also verified
experimentally for arrays of optical waveguides~\cite{exp_mor}.
Defects may provide an additional physical mechanism for light
confinement, and they can support both linear and nonlinear
localized modes, which has been studied theoretically for
different nonlinear models~\cite{kiv,sukh,exp_chen} and observed
experimentally in one-dimensional photonic lattices~\cite{chen2}.

In this Letter, we introduce a novel type of nonlinear defect in
waveguide arrays closely linked with the recently discussed
phase-slip defects in two-dimensional photonic
crystals~\cite{raikh,noda}. In particular, we demonstrate that two
semi-infinite mismatched identical arrays of optical waveguides can
support a variety of linear and nonlinear localized modes with the
specific properties of both discrete solitons~\cite{review} and
nonlinear surface modes~\cite{surface,OL_our}. We analyze stability
of the localized modes and their generation in both linear and
nonlinear regimes.

We consider an array of nonlinear optical waveguides created by two
semi-infinite identical mismatched arrays, as shown in Fig.~1.
In this array, two mismatched waveguides at the sites $n=m$ and
$n=m+1$ interact with a different coupling parameter, $V^{\prime} \neq V$,
so that the coupled-mode system can be described by the
discrete equations for the normalized mode amplitudes $E_n$,
\be i{d E_n\over dz}+V(E_{n+1}+E_{n-1})+ \sum_m V_{nm} E_m +
\gamma |E_{n}|^{2} E_{n}=0, \label{model}
\ee
\be \sum_m V_{nm} E_m = (V'-V)[\delta_{n,m}  E_{m+1}+ \delta_{n,m+1}
E_{m}], \label{model2}
\ee
where $E_n$ are defined in terms of the actual electric fields
${\cal E}_n$ as $E_n=(2\lambda_0\eta_0 /\pi n_0n_2)^{1/2}{\cal
E}_n$, where $\lambda_0$ is the free-space wavelength, $\eta_0$ is
the free-space impedance, $n_2$ and $n_0$ are nonlinear and linear
refractive indices of each waveguide, and $\gamma$ ($\pm 1$) defines
the type of nonlinearity.
%
\begin{figure}[h]
\includegraphics[scale=0.35]{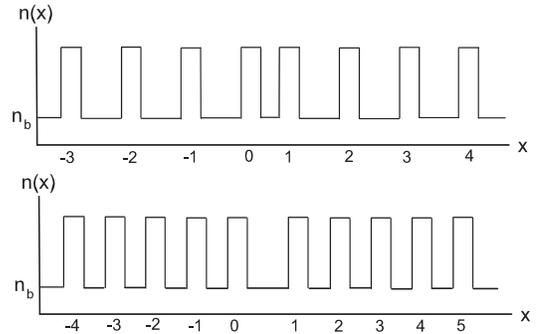}
\caption{Transverse profile of the refractive index for an array
of weakly coupled nonlinear optical waveguides with a phase-slip
defect located between the waveguides at the sites $m=0$ and
$m=1$. Top: $V^{\prime}> V$, bottom: $V^{\prime} <V$.}
\label{fig1}
\end{figure}

We look for stationary solutions  of Eqs.~(\ref{model}),
(\ref{model2}) in the form $E_{n}(z) = E_{n} \exp(i \beta z)$, and
consider first the linear waveguide array (or the limit of low
beam powers) when $\gamma = 0$. In this case, we expect that
localized modes may exist for $V^{\prime}>V$ only, since
decreasing the ratio $V^{\prime}/V$ decouples the chain into two
pieces, and each of the semi-infinite chains does not support
surface modes~\cite{surface,OL_our}. We search for localized
solutions of the form $E_{n} = A \xi^{|n-m|}$, for $n \leq m$, and
$E_n = B \xi^{(n-m-1)}$, for $n\geq m+1$, where $|\xi|<1$. After
some algebra, we obtain $\beta = V [\xi + (1/\xi) ]$, $B/A = \xi
(V^{\prime}/V)$ and $\xi = \pm |V/V^{\prime}|$ so that indeed,
localized modes require the condition $V^{\prime}>V$.

For $\xi = |V/V'|$ and $A=B$ we obtain unstaggered localized modes
[see Fig.~2(a)]: $E_{n}= A |V/V^{\prime}|^{|n-m|}$, for $n \leq
m$, and  $E_n= A|V/V^{\prime}|^{n-m-1}$, for $n\geq m+1$ with the
propagation constant $\beta/V = |V/V^{\prime}| + |V^{\prime}/V|$.
Similarly, for $\xi = -V/V'$ and $A=-B$ we obtain the
corresponding staggered localized modes [see Fig.~2(b)].

\begin{figure}[t]
\noindent\includegraphics[scale=.4]{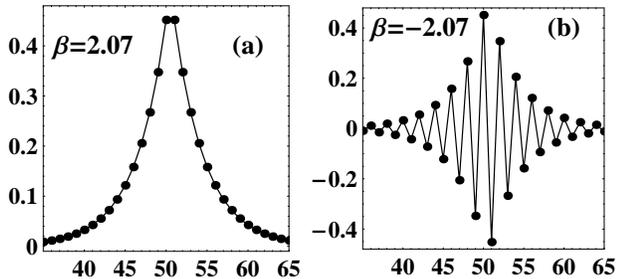} \caption{Linear
unstaggered (a) and staggered (b) localized modes at the
phase-slip defect for $V^{\prime}/V=1.3$ ($N=100,m=50$).}
\label{fig2}
\end{figure}

Next, we consider the nonlinear case described by the stationary
form of Eqs.~(\ref{model}), (\ref{model2}) for $\gamma > 0$. For a
given value of $\beta$, the system of the stationary equations is
solved numerically by a multidimensional Newton-Raphson scheme. As
we are interested in the modes localized near the defect, we look
for the modes with the mode maxima near the slip boundary that
decay quickly away from the bond impurity. We find that the
results vary depending on whether $V^{\prime} > V$ or
$V^{\prime}<V$.
\begin{figure}[h]
\noindent\includegraphics[scale=.5]{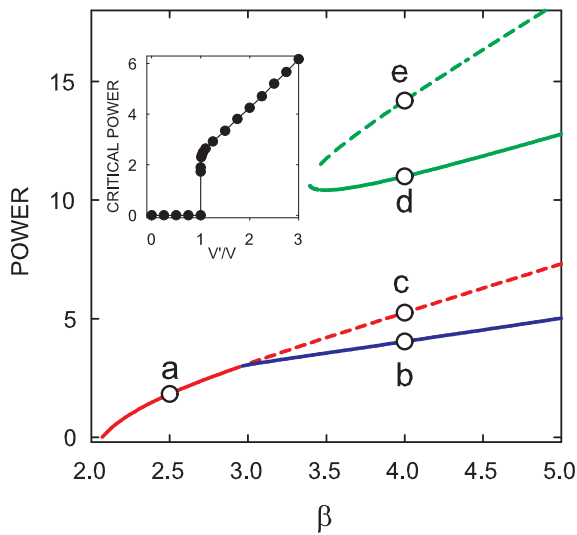} \caption{Power vs.
propagation constant for several families of nonlinear localized
modes at the phase-slip defect for $V^{\prime}/V=1.3$. Solid
(dashed) curves denote stable (unstable) branches. Inset: Minimum
power to destabilize the fundamental mode vs. coupling mismatch.}
\label{fig3}
\end{figure}

\begin{figure}[h]
\noindent\includegraphics[scale=.33]{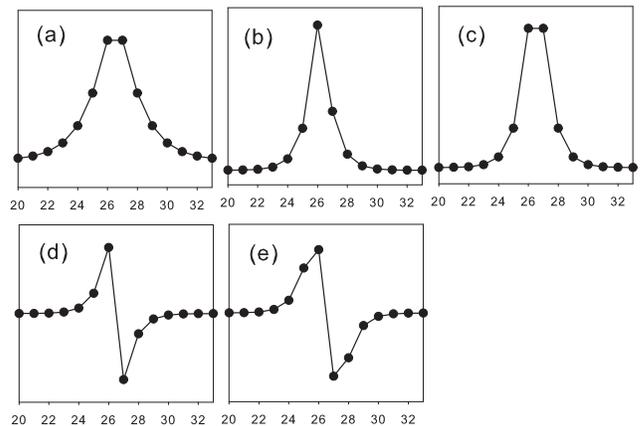} \caption{(a-e)
Examples of the nonlinear localized modes marked by the letters
a,b,c,d, and e in Fig.~\ref{fig3}.} \label{fig4}
\end{figure}

Figures~\ref{fig3} and \ref{fig4} depict the mode power vs.
propagation constant and show specific examples of the mode
profiles, for the case $V^{\prime} > V$. First, one of the
nonlinear modes extends all the way down to zero power, and it
generalizes the linear mode found above [see the curve (a-c) in
Fig.~\ref{fig3} and Figs.~\ref{fig4}(a,c)]. This mode becomes
unstable above a certain threshold power, and it transforms into
an odd mode centered at any of the two equivalent sites coupled by
the bond impurity [see Fig.~\ref{fig4}(b)]. This result can be
easily understood from the known instability of an even mode for a
discrete homogeneous lattice: As the power is increased, the
effective coupling is decreased and the distinction between $V$
and $V^{\prime}$ becomes blurred. In the high-power limit, the
even-like localized state ``feels'' as inside an homogeneous
lattice; hence the onset of instability. The inset in
Fig.~\ref{fig3} shows the critical power needed to destabilize the
fundamental, even-like localized mode. The most interesting
feature of this curve is that it rises very steeply as soon as the
ratio $V^{\prime}/V$ is slightly above one, followed by a slow,
almost linear-like growth. This suggests that a tiny amount of
mismatch is enough to stabilize the even-like mode at the phase
slip.

For larger powers, we find novel types of nonlinear modes
localized at the phase-slip defect [see Figs.~\ref{fig4}(d,e)]
resembling a bound state of two simpler modes. These modes
resemble the so-called twisted modes found earlier in the
homogeneous chain~\cite{sukh,twisted}, and they exist only above a
certain power threshold. The complementary unstable mode looks
like the twisted mode with  ``shoulders'' [See Fig. 4(e)].

\begin{figure}[h]
\noindent\includegraphics[scale=.5]{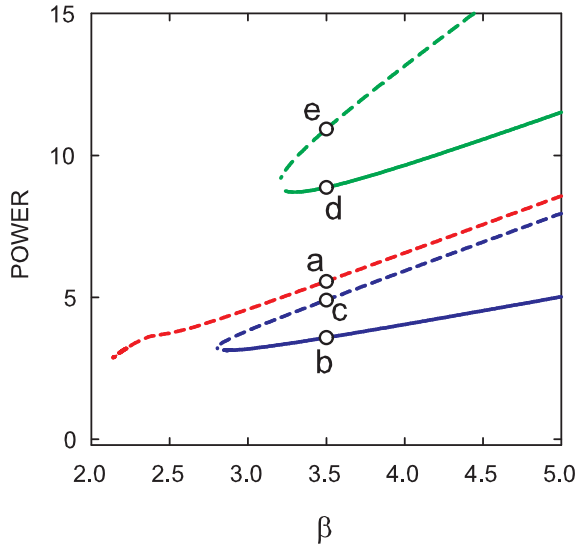} \caption{Power vs.
propagation constant for several localized states at the
phase-slip defect for $V^{\prime}/V=0.7$. Solid (dashed) lines
denote stable (unstable) branches.} \label{fig5}
\end{figure}
\begin{figure}[t]
\noindent\includegraphics[scale=0.6]{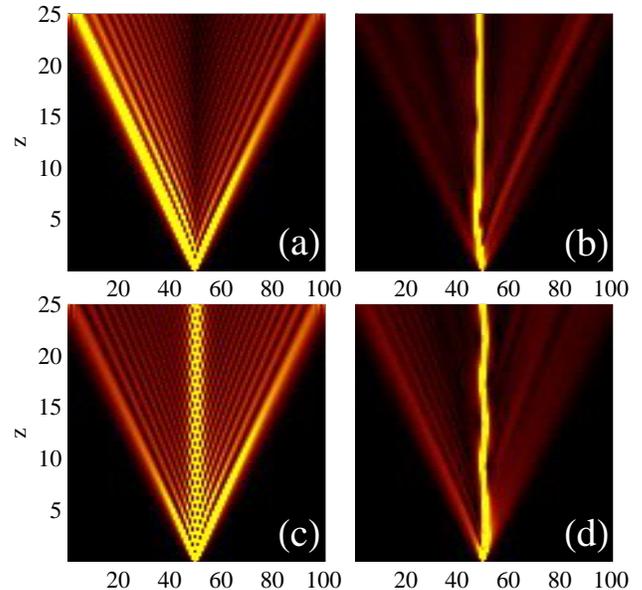} \caption{
One-waveguide excitation of the defect modes. Top: $V^{\prime}/V=
0.7$, i.e. for a weak bond defect, for the input amplitude (a)
$V_0= 1.0$ and (b) $V_0= 2.0$ (b). Bottom: same as above but for
stronger bond defect, $V^{\prime}/V= 1.3$.} \label{fig6}
\end{figure}
In the case of a weaker bond defect, i.e. $V^{\prime} < V$, linear
localized modes do not exist. This result is consistent with the
case of surface modes~\cite{surface,OL_our} where a certain power
threshold is required to support a mode localized at the edge of
the waveguide array. Similarly, in this model the localized mode
appear for finite powers (see the branch marked 'a' in
Fig.~\ref{fig5}). The mode profiles are similar to those presented
in Figs.~\ref{fig4}(a-e). In this case, all localized modes are
strictly nonlinear i.e., they disappear in the limit
$\gamma\rightarrow 0$. As a result, most of those modes are
unstable, and only two modes corresponding to the lower power are
stable.

We should mention that all modes for $\gamma >0$ remain
unstaggered, and staggered modes in this model appear only for
$\gamma=-1$, being found through a simple transformation $E_n
\rightarrow (-1)^n E_n$ applied to all types of modes discussed
above.

Finally, we analyze the generation of the defect modes by an input
beam sent to one of the waveguides of the phase-slip defect. For
weaker coupling ($V^{\prime} < V$) and low powers, we observe no
power localization near the defect sites, and the input power
diffracts as in homogeneous arrays [see Fig.~\ref{fig6}(a)].
However, when we increase the input power, we are able to generate
the asymmetric nonlinear defect mode that correspond to the lowest
branch ( marked with 'b') in Fig.~\ref{fig5} and finite powers,
similar to the excitation of discrete surface
solitons~\cite{surface,OL_our}. On the contrary, the surface mode
is always generated for the case of stronger coupling, $V^{\prime}
> V$, when the defect mode exists in the linear regime, as shown in
Figs.~\ref{fig6}(c,d).

In conclusion, we have introduced and described novel types of
nonlinear defect modes localized at a phase-slip defect in an array
of nonlinear optical waveguides.  We have demonstrated that these
localized modes possess many specific properties of both discrete
solitons and nonlinear surface modes, and they can be easily
generated in both linear and nonlinear regimes.

This work has been supported by Fondecyt grants 1050193 and 7050173
in Chile, and by the Australian Research Council in Australia.
M.I.M. thanks the Nonlinear Physics Center at the Australian
National University for hospitality and financial support.

\end{document}